\documentstyle[12pt,epsfig]{article}
\begin{document}
\baselineskip=23pt
\vspace{1.2cm}
\begin{center}
{\Large \bf Neutron spin polarization in
 strong magnetic fields}
\bigskip
\\
H. Wen$^{1,2}$\footnote{Haibao.Wen@exp2.physik.uni-giessen.de},
L.S.~Kisslinger$^{3}$, Walter~Greiner$^{4}$
and G. Mao$^{2,5,6}$ \\
\bigskip
\bigskip
{\em $^1$ II.Physikalisches Insitut, Justus-Liebig-University Giessen,\\
Heinrich-Buff-Ring 16, D-35392, Giessen, Germany}\\
{\em $^2$ Institute of High Energy Physics, Chinese Academy of Sciences,\\
 P.O.Box 918(4), Beijing 100049, China}\\
{\em $^3$ Department of Physics, Carnegie Mellon University,
Pittsburgh, \\ Pennsylvania 15213, U.S.A.}\\
{\em  $^{4}$ Frankfurt Institute for Advanced Studies,J.W.~Goethe-Universit\"{a}t,\\
Max-von-Laue Strasse 1, 60438 Frankfurt am Main, Germany}\\
 {\em $^5$ CCAST (World Laboratory),P.O.Box 8730, Beijing
100080,China}\\
{\em $^6$ Center of Theoretical Nuclear Physics, National
Laboratory of Heavy Ion Collisions, Lanzhou 730000, China}
\end{center}
\bigskip
\bigskip
\bigskip

\centerline{\large Abstract} The effects of strong magnetic fields
on the inner crust of neutron stars are investigated after taking
into account the anomalous magnetic moments of nucleons. Energy
spectra and wave functions for protons and neutrons in a uniform
magnetic field are provided. The particle spin polarizations and
the yields of protons and neutrons are calculated in a free Fermi
gas model. Obvious spin polarization occurs when $B\geq10^{14}$G
for protons and $B\geq10^{17}$G for neutrons, respectively. It is
shown that the neutron spin polarization depends solely on the
magnetic field strength.


\vspace{1.2cm}
\newpage

\begin{sloppypar}
The observed magnetic field strengths  of neutron stars are in the
range of  $10^{12}\sim10^{14}$G \cite{M91}. For certain magnetars,
which are newly born neutron stars, it can be up to $10^{15}$G
[2-5]. An extra strong magnetic field up to $10^{18}$G \cite{L77}
may exist in the interior of neutron stars according to the
estimation based on the scalar virial theorem (for a neutron star
with radius equaling 10km and mass equaling a sun's). Such strong
field has vast effects on the charged particles as well as
uncharged particles   when the anomalous magnetic moment (AMM)
terms are taken into account. Thus, it influences the structure of
neutron stars. Furthermore, the linear polarization in prompt
$\gamma$-ray emission was discovered for GRB021206 \cite{W03},
which can be attributed to the existence of a strong magnetic
field in fireball. After simple calculation, for a fireball of
$\gamma$-ray burst with radius 100km and total energy
$\displaystyle{2\times10^{51}}$erg one can infer it having a
nuclear matter density about $\displaystyle{10^6}$g/cm$^3$ and a
magnetic field around $\displaystyle{3.5\times10^{15}}$G. The
origin of the magnetic field, though still ambiguous, might be
involved with the surrounding  of fireball sources. The above two
situations contain the similar characteristic, i.e., the presence
of strong magnetic fields in the low density matter.

The effects of the magnetic field enter in two aspects. Firstly,
charged particles in a strong field can be Landau quantized. This
quantization leads to polarization of particles and a consequent
softening of the equation of state (EOS) \cite{C97}. Secondly,
when the AMM term is taken into account, it can cause further
split of energy levels, and the neutral particles can be spin
polarized too. The effects of very strong magnetic fields on the
equation of state for high density matter in neutron stars with
the incorporation of the nuclear AMM have been studied
\cite{B00a,M03c1,M03c2}. An ideal neutron-proton-electron ($npe$)
gas was applied to investigate the properties of low-density
matter in strong magnetic fields, with the incorporation of muon
degree of freedom at high density \cite{S01a}. Nevertheless, the
spin polarization of particles induced by the constant magnetic
field has not yet been discussed thoroughly. It would be
especially interesting for neutral particles since it is simply
the effect of the anomalous magnetic moment terms. In this paper
we investigate the particle spin polarization in a $npe$ system
possessing chemical equilibrium and charge neutrality.
Calculations for pure neutron matter will also be carried out
since illustratively it approaches the situation of inner crust of
neutron stars.

We consider the problem of low-density nuclear matter in an
external magnetic field where the strong interactions are
negligible. The Lagrangian density is written as
\begin{equation}
\label{L1}
\begin{array}{lll}
{\mathcal{L}}&=&
\displaystyle\overline{\psi}[i\gamma_\mu\partial^\mu-e\frac{1+\tau_0}{2}\gamma_\mu
A^\mu-\frac{1}{4}g_b\mu_N\sigma_{\mu\nu}F^{\mu\nu}-m_N]\psi\\[0.5cm]

 &&+\displaystyle\overline{\psi_e}[i\gamma_\mu\partial^\mu-e\gamma_\mu
A^\mu-\frac{1}{4}g_e\mu_B\sigma_{\mu\nu}F^{\mu\nu}-m_e]\psi_e~,
\end{array}
\end{equation}
where $\displaystyle{A^\mu=(0,0,Bx,0)}$ refers to a constant
external magnetic field along $z$ direction,
$\displaystyle{\sigma_{\mu\nu}=\frac{\textit{i}}{2}[\gamma_\mu,\gamma_\nu]}$,
$\displaystyle{F_{\mu\nu}=\partial_\mu{A_\nu}-\partial_\nu{A_\mu}}$,
and $\displaystyle{\tau_0}$ is the third component of the isospin
operator of the nucleon. $\displaystyle{m_N}$ and
$\displaystyle{m_e}$ are the free nucleon mass and electron mass,
$\displaystyle{\mu_N}$ and $\displaystyle{\mu_B}$ are the nuclear
magneton of nucleons and Bohr magneton of electrons,
$\displaystyle{g_p=3.58569470156}$,
$\displaystyle{g_n=-3.8260854690}$, and
$\displaystyle{g_e=\alpha/\pi}$ are the coefficients of the AMM
terms for protons, neutrons and electrons, respectively. Since we
will consider the magnetic field up to $10^{18}$ G, which is
stronger than the critical field of electrons, the higher order
terms of the electron self-energy beyond the AMM term will play a
role. The effects of those high-order terms are sophisticated and
largely cancel the effects of the electron AMM term \cite{D02a}.
Thus, following Ref. \cite{B00a}, for electrons only the
electromagnetic interactions are taken into account.

The Dirac equation for nucleons in a homogeneous magnetic field
can be written as
\begin{equation}
\label{D1} [i\gamma_\mu\partial^\mu-e\frac{1+\tau_0}{2}\gamma_\mu
A^\mu-\frac{1}{4}g_b\mu_N\sigma_{\mu\nu}F^{\mu\nu}-m_N]\psi=0~.
\end{equation}
Solving the above equation in the chiral representation we obtain
the energy spectra and corresponding eigenfunctions. For positive-
and negative-energy protons,
\begin{equation}
\label{pes}
\begin{array}{lll}
\displaystyle{E_p(S,sn,\nu)=sn\left[\left(\sqrt{m_p^2+2eB\nu}+S\Delta_p\right)^2+p_z^2\right]^{1/2}}~,
\\
\end{array}
\end{equation}
\begin{equation}
\label{pee}
\begin{array}{lll}
\displaystyle{\Psi_p(S,sn,\nu)=\sqrt{\frac{E_p(S,sn,\nu)+p_z}{2E_p(S,sn,\nu)}}\frac{\sqrt{2eB\nu}e^{-iE_p(S,sn,\nu)t+ip_yy+ip_zz}}{\left[\left(m_p+S\sqrt{m_p^2+2eB\nu}\right)^2+2eB\nu\right]^{1/2}}}\\[0.5cm]
\\
~~~~~~~~~~~~~~~~~~\times\left(
\begin{array}{cccc}
\displaystyle\frac{i}{\sqrt{2eB\nu}}(m_p+S\sqrt{m_p^2+2eB\nu})I_{\nu;p_y}(x)\\
\\
\displaystyle\frac{E_p(S,sn,\nu)-p_z}{\Delta_p+S\sqrt{m_p^2+2eB\nu}}I_{\nu-1;p_y}(x)\\
\\
\displaystyle\frac{i}{\sqrt{2eB\nu}}\frac{m_p+S\sqrt{m_p^2+2eB\nu}}{\Delta_p+S\sqrt{m_p^2+2eB\nu}}\left(p_z-E_p(S,sn,\nu)\right)I_{\nu;p_y}(x)\\
\\
\displaystyle{I_{\nu-1;p_y}(x)}
\end{array}
\right)~,
\end{array}
\end{equation}
where the Hermite polynomials are defined as given in Ref.
\cite{K83}
\begin{equation}
\label{uf}
\begin{array}{lll}
\displaystyle
I_{\nu;p_y}(x)=\left(\frac{eB}{\pi}\right)^{\frac{1}{4}}
{\mathrm{exp}}\left[-\frac{1}{2}eB\left(x-\frac{p_y}{eB}\right)^2\right]
\frac{1}{\sqrt{\nu!}}H_{\nu}\left[\sqrt{2eB}\left(x-\frac{p_y}{eB}\right)\right]~,\\
\displaystyle
H_{\nu}(x)=(-1)^{\nu}{\mathrm{exp}}\left(\frac{x^2}{2}\right)
\frac{d^{\nu}}{dx^{\nu}}{\mathrm{exp}}\left(-\frac{x^2}{2}\right)~,
~~{\mathrm{for}}~~ \nu=0,1,2,\cdot\cdot\cdot~.
\end{array}
\end{equation}
In the above formulae,
$\displaystyle{\Delta_p=-\frac{1}{2}g_p\mu_NB}$ is the AMM term of
protons, $\displaystyle{\displaystyle{sn=\pm1}}$ denotes the
positive-energy and negative-energy solutions and
$\displaystyle{S=\pm1}$ indicates the spin-up and spin-down
particles, respectively. Here the  spin-up and spin-down are just
relative notions because now the spin operators of considered
particles do not commute with the corresponding Hamiltonian. The
set of functions $I_{\nu;p_y}(x)$ is complete and orthonormal
\cite{K83}, and $\nu$ is the quantum number of Landau levels for
charged particles \cite{L77o,Gre01}. When $\nu=0$, the
eigenfunctions turn out to be
\begin{equation} \label{foup}
\begin{array}{l}
\displaystyle\Psi_p(+1,sn,0)=\sqrt{\frac{E_p(+1,sn,0)+p_z}{2E_p(+1,sn,0)}}e^{-iE_p(+1,sn,0)t+ip_yy+ip_zz}\\[0.5cm]
\\
~~~~~~~~~~~~~~~~~~\times iI_{0;p_y}(x)\left(
\begin{array}{cccc}
1\\
\\
0\\
\\
\displaystyle\frac{p_z-E_p(+1,sn,0)}{\Delta_p+m_p}\\
\\
0
\end{array}
\right)~,
\\
\end{array}
\end{equation}
\begin{equation}
\label{fodp} \displaystyle{\Psi_p(-1,sn,0)=0}~.
\end{equation}
Since neutrons have no charge, they interact with the external
magnetic field merely through the AMM term. The spectra and
eigenfunctions of positive- and negative-energy neutrons read as
\begin{equation}
\label{nes}
\begin{array}{lll}
\displaystyle{E_n(S,sn)=sn\left[\left(\sqrt{m_n^2+p_x^2+p_y^2}+S\Delta_n\right)^2+p_z^2\right]^{1/2}}~,
\\
\end{array}
\end{equation}
\begin{equation}
\label{nee}
\begin{array}{lll}
\displaystyle{\Psi_n(S,sn)=\sqrt{\frac{E_n(S,sn)+p_z}{2E_n(S,sn)}}\frac{\sqrt{p_x^2+p_y^2}e^{-iE_n(S,sn)t+i\overrightarrow{p}\cdot\overrightarrow{x}}}{\left[\left(m_n+S\sqrt{m_n^2+p_x^2+p_y^2}\right)^2+p_x^2+p_y^2\right]^{1/2}}}\\
\\
~~~~~~~~~~~~~~~~~~\times\left(
\begin{array}{cccc}
\displaystyle\frac{m_n+S\sqrt{m_n^2+p_x^2+p_y^2}}{p_x+ip_y}\\
\\
\displaystyle\frac{\Delta_n+S\sqrt{m_n^2+p_x^2+p_y^2}}{p_z+E_n(S,sn)}\\
\\
\displaystyle\frac{p_z-E_n(S,sn)}{p_x+ip_y}\frac{m_n+S\sqrt{m_n^2+p_x^2+p_y^2}}{\Delta_n+S\sqrt{m_n^2+p_x^2+p_y^2}}\\
\\
\displaystyle{1}
\end{array}
\right)~,
\end{array}
\end{equation}
where $\displaystyle{\Delta_n=-\frac{1}{2}g_n\mu_NB}$ is the AMM
term of neutrons. The solutions of Eq. (\ref{D1}) has been given
in Ref. \cite{B00a} in the Dirac representation. The above
formulae obtained in the chiral representation are concise and
compact. One can see that after including the AMM terms the energy
levels of nucleons become spin non-degenerate. Consequently, in
the derivation the wave functions are solely determined without
the need of introducing additional physical quantity to form
common eigenstates. If the AMM term of electrons is taken into
account, one  gets the same energy spectra and eigenfunctions as
those of protons. The solution of electrons without the AMM is
available in Ref. \cite{K83}. We will directly use their results
in the present investigation. The
 electron energy spectra read
\begin{equation}
\label{ees}
\begin{array}{lll}
\displaystyle{E_e(sn,\nu)=sn\sqrt{m_e^2+p_z^2+2eB\nu}}~.
\end{array}
\end{equation}

In a free Fermi gas model the chemical equilibrium is realized
between nucleons and electrons
\begin{equation}
\label{ce}
\begin{array}{lll}
\mu_n=\mu_p+\mu_e~,
\end{array}
\end{equation}
where the chemical potentials are defined as
$\displaystyle{\mu_p=\epsilon_p^f}$,
$\displaystyle{\mu_n=\epsilon_n^f}$ and
$\displaystyle{\mu_e=\epsilon_e^f}$. They are related with Fermi
momenta as follows
\begin{equation}
\label{pFmFe}
\begin{array}{lll}
\left[k_p^f(S,\nu)\right]^2=\left(\epsilon_p^f\right)^2-\left(\sqrt{m_p^2+2eB\nu}+S\Delta_p\right)^2~,
\end{array}
\end{equation}
\begin{equation}
\label{nFmFe}
\begin{array}{lll}
\left[k_n^f(S)\right]^2=\left(\epsilon_n^f\right)^2-\left(m_n+S\Delta_n\right)^2~,
\end{array}
\end{equation}
\begin{equation}
\label{eFmFe}
\begin{array}{lll}
\left[k_e^f(\nu)\right]^2=\left(\epsilon_e^f\right)^2-\left(m_e^2+2eB\nu\right)~,
\end{array}
\end{equation}
which are determined through respective particle number densities
\begin{equation}
\label{pndFm}
\begin{array}{lll}
\displaystyle{\rho_p^0=\frac{eB}{2\pi^2}\sum_{S}\sum_{\nu}k_p^f(S,\nu)}~,
\end{array}
\end{equation}
\begin{equation}
\label{nndFm}
\begin{array}{lll}
\displaystyle\rho_n^0=\frac{1}{2\pi^2}\sum_{S}\left\{\frac{1}{3}\left[k_n^f(S)\right]^3+\frac{S\Delta_n}{2}\right.\\[0.5cm]
\left.\displaystyle
~~~~~~\times\left[(m_n+S\Delta_n)k_n^f(S)+(\epsilon_n^f)^2\left(\arcsin\frac{m_n+S\Delta_n}{\epsilon_n^f}-\frac{\pi}{2}\right)\right]\right\}~,
\end{array}
\end{equation}
\begin{equation}
\label{endFm}
\begin{array}{lll}
\displaystyle{\rho_e^0=\frac{eB}{2\pi^2}\sum_{S}\sum_{\nu}k_e^f(\nu)}~.
\end{array}
\end{equation}
The index $\displaystyle{\nu}$ runs up to the largest integer for
$\displaystyle{[k_p^f(S,\nu)]^2}$ to be positive according to Eq.
(\ref{pFmFe}). $\rho_p^0=\rho_e^0$ is required for charge neutral
matter. We consider the density from neutron drip point
($4.3\times10^{11}\mathrm{g/cm^3}$) \cite{B71a} to a quarter of
nuclear saturation density, i.e., around
$10^{11}\sim10^{14}\mathrm{g/cm^3}$. The nonlinear equations
preserving the charge neutral and chemical equilibrium \cite{G97n}
conditions are solved numerically in an iteration procedure.

\begin{figure}[h]
\centerline{
\includegraphics[width=4in]{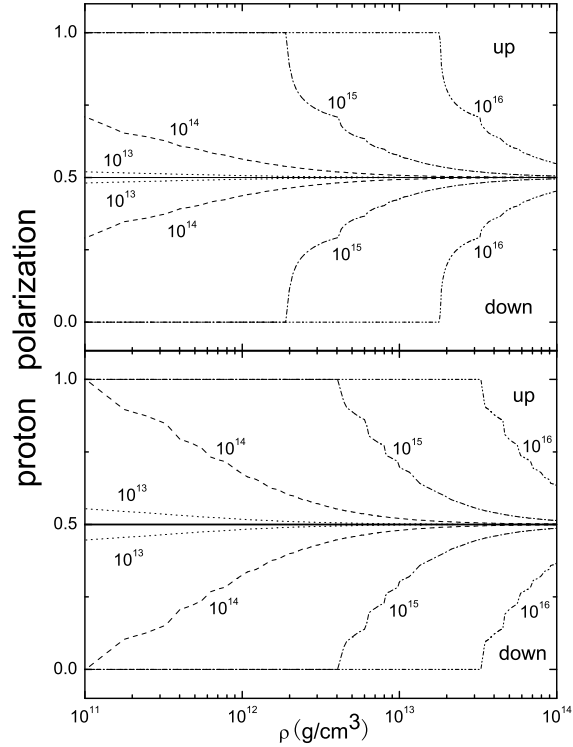}}
\caption{Proton spin polarization at different magnetic field
strengths as a function of density. The upper and lower panel
represent the conditions of without and with the AMM term of
protons. The magnetic field strengths are tagged on the lines.}
\label{pp}
\end{figure}

Figure~\ref{pp} shows the spin ratio of protons in a \textit{npe}
system as a function of density. Different magnetic field
strengths are considered in calculations as indicated in the
figure. We see that the proton spin polarization increases with
the increasing of the magnetic field strength and the decreasing
of baryon density. The incorporating of the AMM term benefits the
spin polarization. Evident proton spin polarization is exhibited
at $B\geq10^{14}$G. On the other hand, protons become completely
spin polarized throughout the range of density when
$B\geq10^{17}$G.

\begin{figure}[h]
\centerline{
\includegraphics[width=4in]{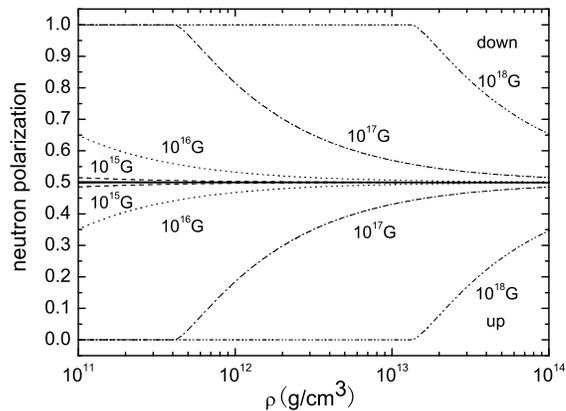}}
\caption{Neutron spin polarization at different magnetic field
strengths as a function of density. The figure displays the
results of pure neutron matter and chemical-equilibrated $npe$
system.} \label{np}
\end{figure}

The spin ratios of neutrons in a \textit{npe} system with the AMM
term incorporated are reckoned at different magnetic field
strengths and displayed in figure~\ref{np} as a function of
density. We can see that the neutron spin polarization increases
with the increasing of the magnetic field strength and the
decreasing of baryon density, too. However, the neutron
polarization is much weaker than that of protons because it is
simply caused by the AMM term. Due to the opposite sign of this
term neutrons incline to occupy the spin-down direction. Complete
neutron spin polarization appears at B $\sim$ 10$^{17}$ G. We have
also investigated the pure neutron matter. It is interesting to
see that the resultant figure of neutron spin polarization is the
same as given in Fig.~2, while in the $npe$ system the particle
fraction changes substantially. The ratios of proton number to the
nucleon number are depicted in Fig.~3 for chemical-equilibrated
$npe$ system. At strong magnetic field the proton fraction is
quite large. The same neutron spin polarization in pure neutron
matter and $npe$ matter indicates that it is sensitive to the
magnetic field strength only.

In summary, the Dirac equations of protons and neutrons are solved
in the chiral representation for a constant magnetic field with
the anomalous magnetic moment terms taken into account. The
properties of low-density matter in the presence of strong
magnetic fields are investigated in a relativistic Fermi gas
model. Obvious spin polarizations for protons and neutrons are
obtained. It turns out that the neutron spin polarization is the
same both in the pure neutron matter and the chemical-equilibrated
$npe$ system.

\textbf{Acknowledgements:} This work was supported by the National
Natural Science Foundation of China under Grant No. 10275072.

\end{sloppypar}

\begin{figure}[h]
\centerline{
\includegraphics[width=4in]{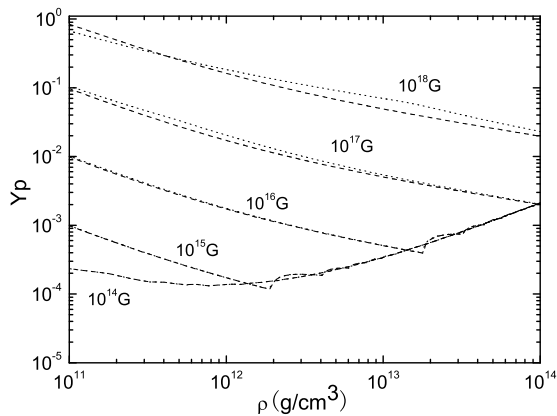}}
\caption{Proton fraction calculated at different magnetic field
strengths as a function of density. The cases without and with the
AMM term are denoted by dashed and dotted lines, respectively.}
\label{pr}
\end{figure}

\end{document}